\documentclass{elsevier-book}

\usepackage{graphicx,url}
\usepackage{array,booktabs,threeparttable,longtable}
\usepackage{cite}
\usepackage{enumitem}
\usepackage{algorithm,algorithmic}

\usepackage{listings}
\lstdefinelanguage{JavaScript}{
  keywords={break, case, catch, continue, debugger, default, delete, do, else, false, finally, for, function, if, in, instanceof, new, null, return, switch, this, throw, true, try, typeof, var, void, while, with},
  morecomment=[l]{//},
  morecomment=[s]{/*}{*/},
  morestring=[b]',
  morestring=[b]",
  sensitive=true
}
\usepackage{amsmath,amssymb}

\usepackage{amsthm}

\theoremstyle{remark}

\begin{document}
\chapter{Blockchain-Enabled Federated Learning}

\chapterauthors{Murtaza Rangwala$^1$, K.R. Venugopal$^2$, and Rajkumar Buyya$^1$}{%
$^1$Quantum Cloud and Distributed Systems (qCLOUDS) Lab, School of Computing and Information Systems, The University of Melbourne, Australia\\[0.3em]
$^2$Department of Computer Science and Engineering, University of Visvesvaraya College of Engineering, Bangalore University, India}

\begin{abstract}
Blockchain-enabled federated learning (BCFL) addresses fundamental challenges of trust, privacy, and coordination in collaborative AI systems. This chapter provides comprehensive architectural analysis of BCFL systems through a systematic four-dimensional taxonomy examining coordination structures, consensus mechanisms, storage architectures, and trust models. We analyze design patterns from blockchain-verified centralized coordination to fully decentralized peer-to-peer networks, evaluating trade-offs in scalability, security, and performance. Through detailed examination of consensus mechanisms designed for federated learning contexts, including Proof of Quality and Proof of Federated Learning, we demonstrate how computational work can be repurposed from arbitrary cryptographic puzzles to productive machine learning tasks. The chapter addresses critical storage challenges by examining multi-tier architectures that balance blockchain's transaction constraints with neural networks' large parameter requirements while maintaining cryptographic integrity. A technical case study of the TrustMesh framework illustrates practical implementation considerations in BCFL systems through distributed image classification training, demonstrating effective collaborative learning across IoT devices with highly non-IID data distributions while maintaining complete transparency and fault tolerance. Analysis of real-world deployments across healthcare consortiums, financial services, and IoT security applications validates the practical viability of BCFL systems, achieving performance comparable to centralized approaches while providing enhanced security guarantees and enabling new models of trustless collaborative intelligence.
\end{abstract}


\section{Introduction}

\subsection{The Collaborative Learning Challenge}

Modern machine learning faces a fundamental paradox: the most valuable models require diverse large-scale datasets, yet privacy regulations, competitive concerns, and data sovereignty requirements increasingly prevent organizations from sharing the data that could significantly improve these models~\cite{mcmahan2017communication}. 

Consider healthcare consortiums attempting to develop diagnostic machine learning systems. Although each participating hospital possesses unique patient data that could collectively improve diagnostic accuracy, HIPAA regulations prohibit data sharing for purposes not related to treatment, payment, or healthcare operations~\cite{tariq2023patient}. Federated learning emerges as a promising solution to this dilemma, enabling collaborative model training without requiring data movement~\cite{kairouz2021advances}. However, federated learning systems introduce significant coordination challenges: Participants must establish mutual trust, verify model integrity across distributed training rounds, assess contributions fairly, and maintain resilience against malicious actors~\cite{mothukuri2021survey}. 

\subsection{Blockchain as the Trust Infrastructure}

These coordination challenges require mechanisms to establish trust and maintain accountability between distributed participants. Blockchain technology addresses this requirement by providing a decentralized infrastructure that can operate with varying levels of centralization based on specific trust assumptions~\cite{li2022blockchain,nguyen2021federated}. Through consensus protocols—particularly Byzantine fault-tolerant mechanisms—blockchain systems ensure transparent, tamper-proof records of all interactions and resilience against malicious behavior~\cite{castro1999practical}. Smart contracts further enhance accountability by automating fair reward distribution based on verified contributions~\cite{kang2019incentive}.

This integration has led to the emergence of a distinct class of \emph{blockchain-enabled federated learning} (BCFL) systems that combine federated learning's privacy preservation with blockchain's verifiability and auditability. BCFL systems solve the collaborative learning paradox by enabling secure and accountable cooperation without compromising data sovereignty, with varying architectural choices supporting different trust models and operational requirements.

\subsection{Chapter Organization}

This chapter systematically analyzes BCFL systems through a comprehensive taxonomy that encompasses four fundamental dimensions. We maintain this taxonomic structure throughout the chapter, ensuring each design dimension receives thorough technical analysis with practical implementation considerations.

A technical case study of the \textsc{TrustMesh} framework~\cite{rangwala2025trustmesh} illustrates how these taxonomic dimensions manifest in real implementations and demonstrates practical BCFL deployment through a simple distributed image classification scenario. We then explore current challenges and future research directions before concluding with key insights and practical guidance for BCFL system design.

\section{A Taxonomy of BCFL Architectures}

\begin{figure*}[!t]
    \centering
    \includegraphics[width=1\textwidth]{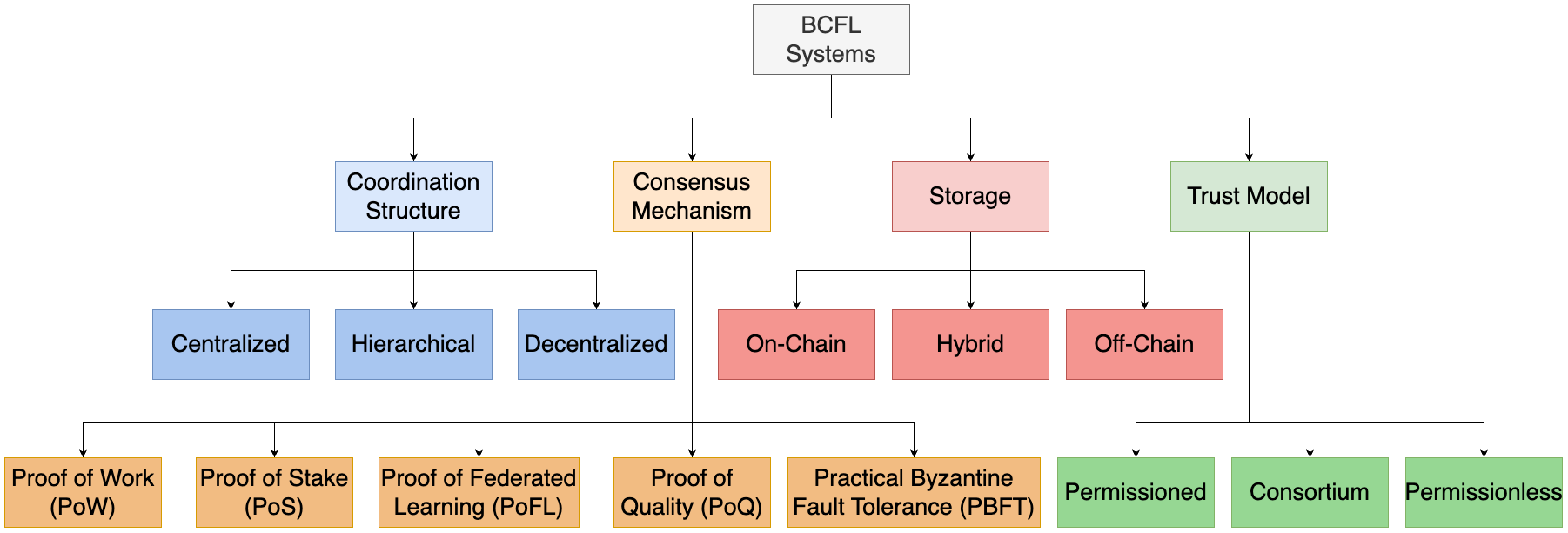}
    \caption{Classification framework for BCFL systems based on coordination, consensus, storage, and trust dimensions. Each dimension offers multiple implementation choices that collectively determine system capabilities and constraints.}
    \label{fig:bcfl-taxonomy}
\end{figure*}

The design space for BCFL systems encompasses numerous architectural choices, each with distinct implications for performance, security, and scalability. Our taxonomy, illustrated in Figure~\ref{fig:bcfl-taxonomy}, systematically categorizes these design alternatives in four fundamental dimensions that collectively define the characteristics and capabilities of the system:

\begin{itemize}
    \item \textbf{Coordination Structure}: Determines the interaction patterns between participants during federated learning rounds. The coordination structure analysis explores centralized, hierarchical, and decentralized approaches with their respective trade-offs between scalability, control, and implementation complexity.
    
    \item \textbf{Consensus Mechanism}: Defines the process by which the network reaches agreement on model updates and maintains system integrity. Consensus mechanisms specifically adapted for federated learning contexts demonstrate traditional blockchain protocols evolving to meet distributed machine learning demands, moving beyond energy-intensive mining to quality-based selection.
    
    \item \textbf{Storage Strategy}: Specifies the location and persistence methods for model parameters, gradients, and metadata. Storage strategies determine data placement and access patterns that influence both system performance and security guarantees.
    
    \item \textbf{Trust Model}: Establishes the security assumptions about participant behavior and potential adversarial actions. Trust models define participation requirements and access control mechanisms, ranging from permissionless open participation to consortium-based collaboration and fully permissioned systems.
\end{itemize}

Each dimension offers multiple implementation choices that can be combined to create BCFL systems tailored to specific requirements. The following sections examine each dimension in detail, analyzing the technical trade-offs and practical considerations that guide architectural decisions.

\section{Coordination Structure}

The coordination structure dimension of our taxonomy examines how participants coordinate during federated learning rounds. Three primary coordination structures have emerged from recent BCFL implementations, each representing different trade-offs between decentralization, scalability, and implementation complexity: centralized coordination with blockchain verification, hierarchical multi-layer architectures, and decentralized peer-to-peer networks.

\subsection{Centralized Coordination with Blockchain Verification}

The most straightforward BCFL architecture preserves traditional federated learning's star topology while incorporating blockchain for verification and immutable audit trails. This approach allows organizations to leverage existing federated learning infrastructure while gaining the trust and transparency benefits of blockchain technology.

\begin{figure*}[!t]
    \centering
    \includegraphics[width=1\textwidth]{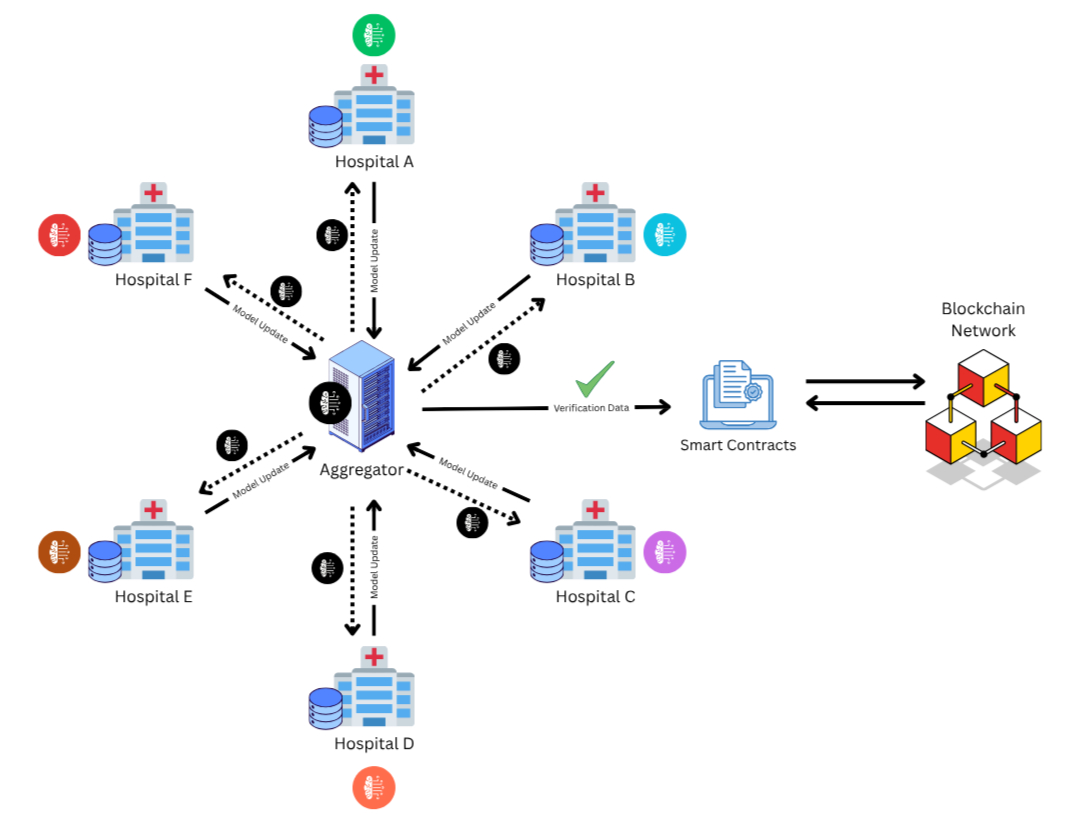}
    \caption{Centralized BCFL architecture showing traditional federated learning coordination enhanced with blockchain verification and smart contract automation. The central aggregator coordinates training while blockchain provides immutable audit trails.}
    \label{fig:centralized-bcfl}
\end{figure*}

As shown in Figure~\ref{fig:centralized-bcfl}, the architecture integrates four essential components in a hub-and-spoke model: a central aggregator that manages training rounds and performs model aggregation, a blockchain network that operates parallel to the learning process for immutable verification, smart contracts that automate governance functions, including participant verification and contribution tracking, and participating organizations that maintain complete control over their local data. The coordination process follows a structured pattern in which participants receive the global model from the central aggregator, train locally on private datasets, and submit encrypted model updates directly to the aggregator. The aggregator performs model aggregation while simultaneously coordinating with the blockchain network for verification and audit purposes, interacting with smart contracts to verify participant authenticity, track contributions, and manage reward distribution.

This coordination structure achieves $\mathcal{O}(n)$ communication complexity for federated learning coordination, where $n$ represents the number of participants, matching traditional federated learning approaches. The blockchain consensus process adds overhead ranging from $\mathcal{O}(b^2)$ for PBFT-style protocols~\cite{castro1999practical} to $\mathcal{O}(b)$ for simpler voting schemes, where $b$ is the number of blockchain nodes. However, this overhead is isolated to the aggregator's interaction with the blockchain network rather than involving all participants. 

The centralized coordination model proves particularly suitable for consortium settings where participants have established trust relationships and regulatory requirements mandate comprehensive audit trails. Recent implementations have validated this approach's practical viability. The BIT-FL framework~\cite{ying2024bitfl} demonstrated how centralized coordination can be enhanced with Byzantine fault tolerance and differential privacy verification while preserving learning performance with minimal coordination overhead. Similarly, FLchain~\cite{majeed2019flchain} leveraged Merkle Patricia Trees to track global model evolution while maintaining the coordination patterns of centralized aggregation, demonstrating that blockchain integration can preserve key coordination advantages through careful architectural design.

\subsection{Hierarchical Multi-Layer Architectures}

\begin{figure*}[ht]
    \centering
    \includegraphics[width=1\textwidth]{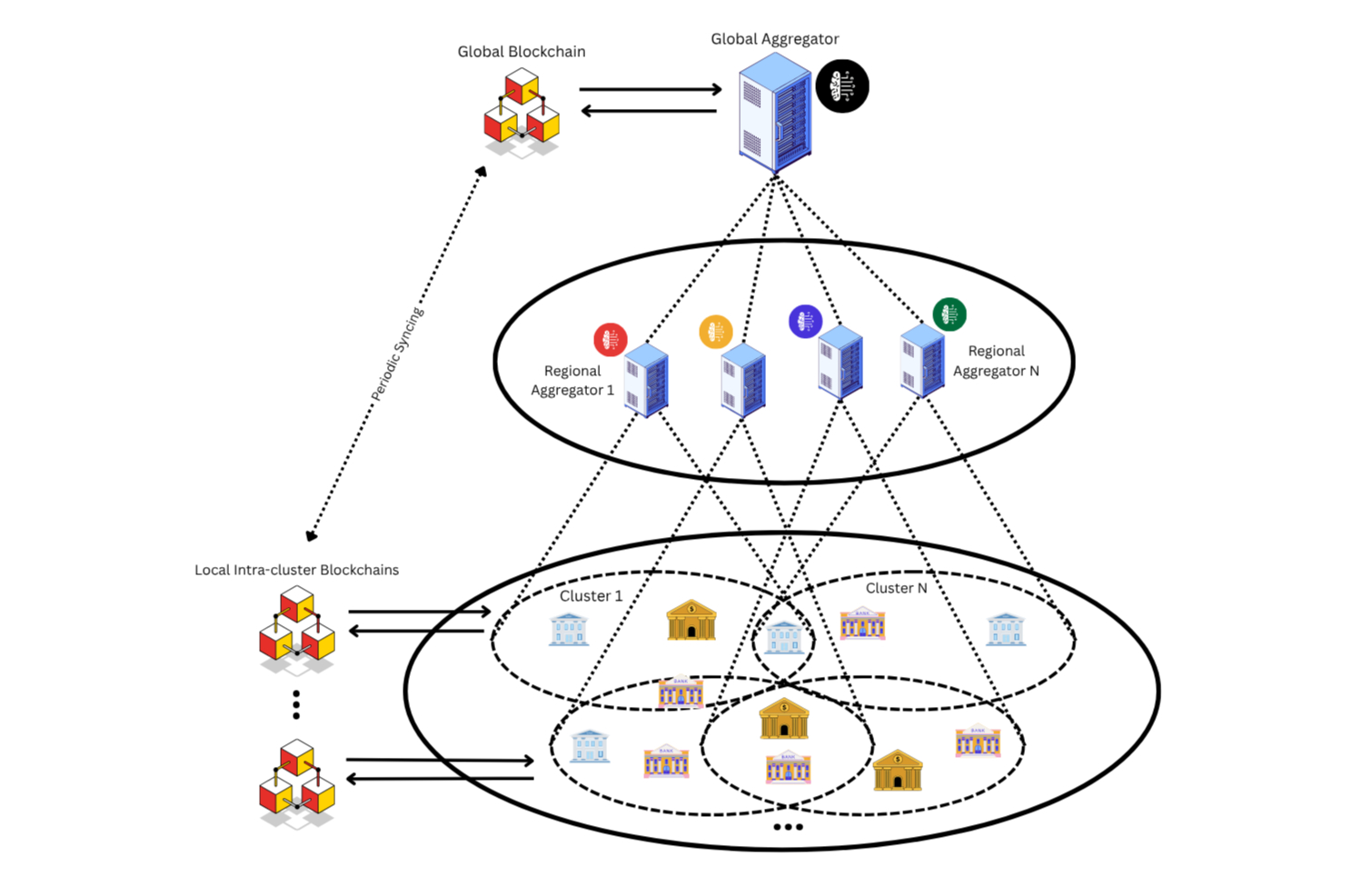}
    \caption{Hierarchical BCFL architecture demonstrating multi-layer aggregation with regional coordination and global consensus. Regional aggregators reduce communication overhead while maintaining global model consistency through cross-layer synchronization protocols.}
    \label{fig:hierarchical-bcfl}
\end{figure*}

While centralized coordination provides familiar patterns and straightforward implementation, it faces inherent scalability limitations as the number of participants grows. Hierarchical BCFL addresses these scalability challenges by organizing participants into multiple coordination layers that mirror natural organizational and geographical boundaries~\cite{chai2020hierarchical}. This structure, depicted in Figure~\ref{fig:hierarchical-bcfl}, recognizes that many real-world federated learning scenarios involve natural hierarchies and leverages these relationships for efficient coordination.

Hierarchical BCFL systems organize participants across multiple coordination layers, with each layer optimized for specific communication patterns and trust relationships. At the local level, participants coordinate within clusters based on geography, institutional affiliation, or data characteristics. Intermediate layers manage regional coordination by collecting and pre-aggregating updates from local clusters, while the global layer maintains system-wide consistency and distributes final model updates across all regions. This layered approach can reduce communication complexity from $\mathcal{O}(n)$ to $\mathcal{O}(\log n)$ in well-balanced hierarchies by limiting direct global communication to regional representatives rather than requiring all participants to communicate globally.

Blockchain integration in hierarchical BCFL systems is usually implemented through several generalizable architectural patterns. A unified blockchain network can record all coordination activities across the hierarchy, providing complete audit trails and simplified consensus mechanisms, though this approach faces scalability limitations as the number of hierarchical layers increases. Alternatively, separate blockchain networks can operate at different hierarchical levels with specialized consensus mechanisms optimized for each tier's requirements—local blockchains handle high-frequency coordination within clusters, while higher-tier blockchains manage cross-regional consensus. Sidechain-based coordination allows regional coordinators to operate semi-independent networks anchored to a main blockchain for final settlement, enabling high-throughput local coordination while maintaining global security guarantees. Hybrid patterns employ blockchain selectively for critical security functions such as participant authentication and final model verification, while traditional coordination mechanisms handle routine tasks.

The effectiveness of hierarchical coordination has been demonstrated across diverse domains, with frameworks showing both theoretical scalability advantages and practical implementation benefits. Chai et al.~\cite{chai2020hierarchical} implemented Internet of Vehicles coordination through a three-layer hierarchy consisting of vehicles, roadside units (RSUs), and base stations (BSs), achieving improved learning accuracy and sharing efficiency compared to conventional federated learning approaches. In IoT security applications, the HBFL framework~\cite{sarhan2022hbfl} successfully deployed edge-fog-cloud hierarchies where edge devices performed local coordination, fog nodes handled intermediate aggregation, and cloud infrastructure managed global consistency. Both frameworks validated the benefits of hierarchical organization and were designed to scale to large IoT deployments through localized processing and reduced communication overhead.

\subsection{Decentralized Peer-to-Peer Networks}

\begin{figure*}[ht]
    \centering
    \includegraphics[width=1\textwidth]{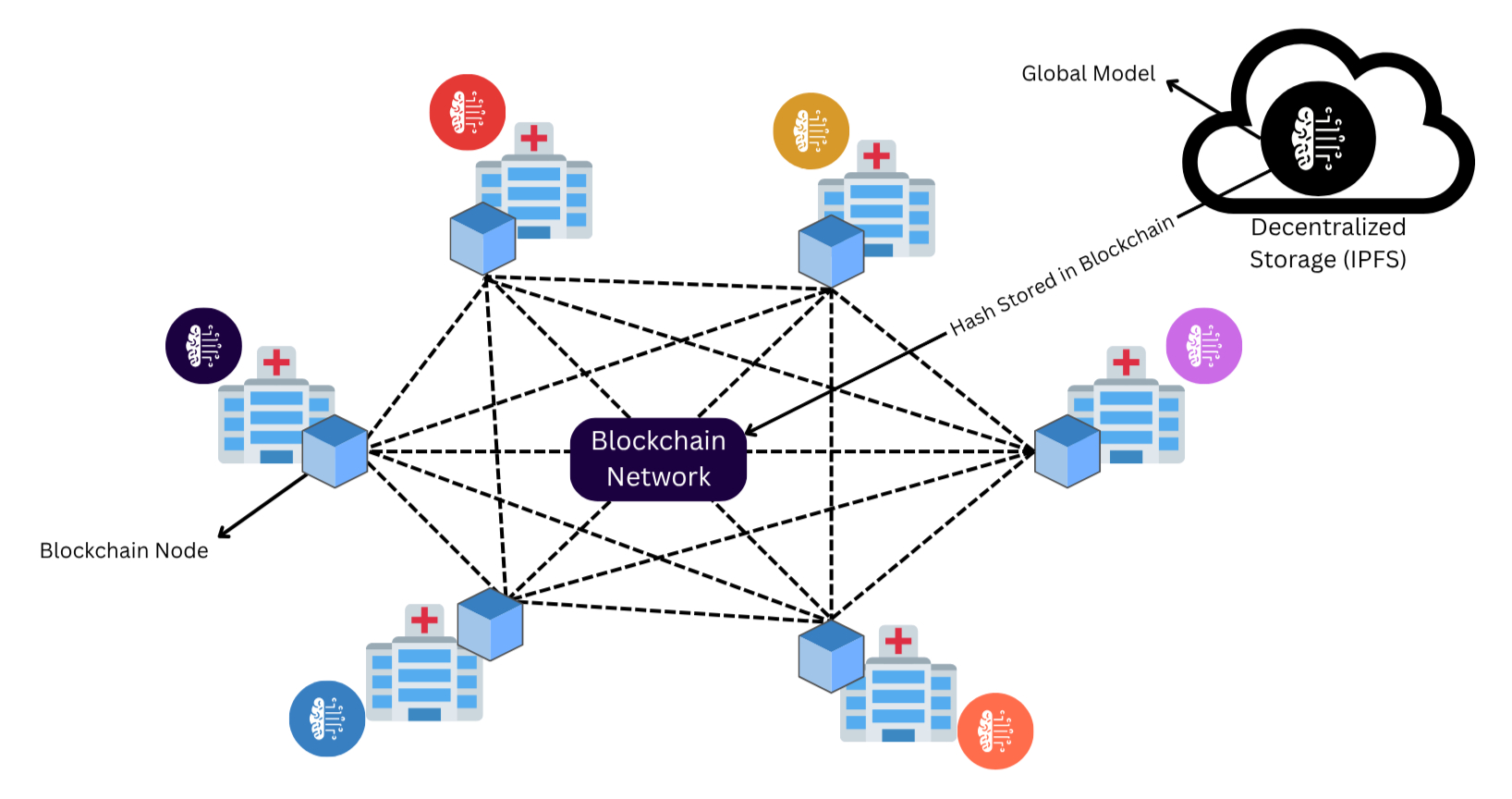}
    \caption{Fully decentralized BCFL architecture showing peer-to-peer coordination through blockchain consensus, with participants conducting distributed model aggregation without central authorities.}
    \label{fig:decentralized-bcfl}
\end{figure*}

Fully decentralized BCFL eliminates all coordination hierarchies, with participants communicating directly through blockchain-mediated distributed protocols~\cite{shayan2020biscotti}. This architecture, illustrated in Figure~\ref{fig:decentralized-bcfl}, represents the most radical departure from traditional federated learning coordination, where every participant functions as both learner and coordinator while the blockchain network serves as the primary coordination infrastructure.

This architecture integrates blockchain as the foundational coordination layer rather than a verification overlay. Participants maintain direct peer-to-peer connections through dynamic communication graphs where the blockchain network orchestrates training rounds and manages consensus. Smart contracts automate participant discovery, training round initiation, and model verification without requiring designated coordinators, with all participants contributing equally to both federated learning and blockchain consensus processes. The coordination process operates through blockchain-mediated consensus where participants query the blockchain for current global model parameters, retrieve models from distributed storage, train locally while forming peer groups for direct model update exchange, then submit aggregated results through blockchain transactions with cryptographic proofs of training correctness.

This coordination structure achieves communication complexity ranging from $\mathcal{O}(n^2)$ for full mesh connectivity to $\mathcal{O}(n \log n)$ with structured overlay networks. While this represents higher overhead than centralized approaches, the elimination of coordination bottlenecks improves system resilience and eliminates single points of failure. The decentralized coordination model proves particularly valuable in scenarios requiring maximum censorship resistance and privacy preservation, where participants cannot trust centralized coordinators. Cross-organizational collaborations between competitors, community-driven learning networks, and applications requiring regulatory independence represent optimal use cases for this architectural approach.

Despite the theoretical complexity challenges, recent implementations have demonstrated practical viability of this coordination model at significant scale. For example, BlockDFL~\cite{qin2024blockdfl} addressed poisoning attacks and Byzantine failures through PBFT-based voting mechanisms and two-layer scoring systems while maintaining complete decentralization, achieving accuracy comparable to centralized federated learning with up to 200 participants. These results validate that decentralized coordination can achieve both the scale and performance necessary for practical deployment.

\begin{table}[ht]
\centering
\begin{threeparttable}
\caption{Coordination structure comparison across key properties}\label{tab:coordination-comparison}
\begin{tabular}{@{}lccc@{}}
\toprule
\textbf{Property} & \textbf{Centralized} & \textbf{Hierarchical} & \textbf{Decentralized} \\
\midrule
Coordination Complexity & $\mathcal{O}(n)$ & $\mathcal{O}(\log n)$ & $\mathcal{O}(n \log n)$ \\
Single Point of Failure & Yes & Partial & No \\
Scalability & Limited (bottleneck) & Very High & High (complex) \\
Implementation Complexity & Low & Medium & High \\
Censorship Resistance & Low & Medium & High \\
Familiar Patterns & High & Medium & Low \\
Organizational Alignment & Consortium & Hierarchical & Peer-to-peer \\
\bottomrule
\end{tabular}
\end{threeparttable}
\end{table}

Table~\ref{tab:coordination-comparison} summarizes the key characteristics and trade-offs across all three coordination structures, providing a framework for selecting the most appropriate approach based on specific deployment requirements and organizational constraints.

\section{Consensus Mechanisms}
\label{sec:consensus}

The consensus mechanism dimension of our taxonomy directly influences all other architectural choices. Consensus protocols determine the validation processes for coordination structures and model updates, dictate storage architectures' approach to data integrity, and define trust models' security guarantee enforcement. Traditional blockchain consensus mechanisms~\cite{nakamoto2008bitcoin,wood2014ethereum} prove suboptimal for federated learning applications due to energy consumption, latency, and misaligned incentives. The unique requirements of BCFL systems—including model quality verification, participant contribution assessment, and privacy preservation—necessitate specialized consensus protocols that align computational work with learning objectives rather than arbitrary cryptographic puzzles.

\subsection{Traditional Consensus Mechanisms in BCFL Context}

Before examining federated learning-specific consensus mechanisms, we establish how traditional blockchain consensus approaches apply to BCFL systems, highlighting their limitations and motivating the need for specialized alternatives.

The traditional Proof of Work (PoW) protocol requires participants to solve computationally expensive cryptographic puzzles to validate transactions and create new blocks. In the BCFL context, PoW presents significant drawbacks, including computational work that does not provide federated learning value, energy consumption that conflicts with collaborative learning sustainability goals, and mining latency that severely affects federated learning round times~\cite{huang2022blockchain}.

To address these computational and energy efficiency concerns, the Proof of Stake (PoS) protocol selects validators based on their stake in the network rather than computational power, offering faster finality and lower energy consumption suitable for resource-constrained participants. However, traditional PoS does not consider the quality of machine learning contributions, potentially allowing participants with large stakes but poor data to dominate consensus. This misalignment between economic power and learning value motivates the development of ML-specific consensus mechanisms.

\subsection{Proof of Quality (PoQ)}

Proof of Quality represents a paradigm shift in consensus design, selecting consensus leaders based on model prediction accuracy rather than computational power or financial stake. This approach directly addresses the critical challenge of model poisoning attacks while ensuring that only high-quality model contributions advance the federated learning process.

This consensus protocol operates through a multi-phase process where participants first train local models and compute validation accuracy using shared validation datasets. Participants then submit cryptographically signed model performance proofs to the blockchain, enabling transparent verification without revealing private training data. The participant with the highest verified accuracy becomes the round leader, responsible for aggregating submitted models and proposing the global update. Other participants verify the aggregation through Byzantine fault-tolerant voting mechanisms before accepting the new global state. Algorithm~\ref{alg:poq-consensus} depicts the complete protocol flow.

\begin{algorithm}
\caption{Proof of Quality Consensus Protocol}
\label{alg:poq-consensus}
\begin{algorithmic}[1]
\STATE \textbf{Input:} Previous global model $w^{(t-1)}$, validation dataset $\mathcal{D}_{val}$
\STATE \textbf{Local Training Phase:}
\FOR{each participant $i$}
    \STATE Train local model: $w_i^{(t)} = \text{LocalTrain}(w^{(t-1)}, \mathcal{D}_i)$
    \STATE Compute validation accuracy: $\text{acc}_i = \frac{1}{|\mathcal{D}_{val}|} \sum_{(x,y) \in \mathcal{D}_{val}} \mathbb{I}[f_i(x) = y]$
    \STATE Generate cryptographic proof: $\text{proof}_i = \text{Sign}(\text{acc}_i, w_i^{(t)})$
    \STATE Submit $(\text{acc}_i, \text{proof}_i)$ to blockchain
\ENDFOR
\STATE \textbf{Leader Selection:} $\text{leader} = \arg\max_i \text{acc}_i$
\STATE \textbf{Aggregation:} Leader aggregates models and proposes global update
\STATE \textbf{Verification:} Network validates aggregation through Byzantine voting
\end{algorithmic}
\end{algorithm}

In this approach, validation dataset management emerges as a critical security consideration. Distributed validation approaches partition validation data across multiple validators using secret sharing schemes, while committee-based validation selects trusted members to evaluate model quality collectively~\cite{chen2021robust}. Security measures include homomorphic encryption for validation datasets, permissioned access control through smart contracts, and Merkle tree structures to ensure dataset integrity. However, sophisticated attack vectors threaten PoQ systems, including validation dataset poisoning where adversaries corrupt validation data to manipulate quality assessments, and model quality spoofing through carefully crafted parameters to artificially inflate scores.

Recent implementations have explored quality-based consensus mechanisms in blockchain-federated learning systems. Lu et al.~\cite{lu2019blockchain} proposed integrating federated learning with permissioned blockchain consensus for industrial IoT data sharing, developing a training quality-based consensus protocol where committee nodes evaluate model quality through Mean Absolute Error (MAE) measurements. Their system demonstrated how consensus leadership could be determined by model performance rather than computational puzzles, showing the feasibility of repurposing federated learning computations for blockchain consensus decisions. Similarly, Chen et al.'s VBFL framework~\cite{chen2021robust} demonstrated remarkable resilience through decentralized validation mechanisms, where individual validators examined the legitimacy of local model updates using validator-threshold based validation combined with Proof-of-Stake inspired consensus. With 15\% malicious devices, VBFL maintained 87\% accuracy on MNIST, representing a 7.4 times improvement over vanilla federated learning under adversarial conditions.

While these quality-based approaches show significant promise for improving model integrity and reducing computational waste compared to traditional PoW systems, they generally still require additional verification and validation work beyond the core federated learning training process. The need for separate validation datasets, cryptographic proof generation, and Byzantine voting mechanisms introduces overhead that, while more efficient than hash-based puzzles, still represents computational work distinct from the productive machine learning task. This limitation motivates the development of consensus mechanisms that more completely integrate productive computation with blockchain security requirements.

\subsection{Proof of Federated Learning (PoFL)}

Proof of Federated Learning ingeniously re-purposes the computational work required for blockchain consensus to perform federated learning training, making consensus work directly productive rather than wasteful. This energy-recycling approach addresses both environmental concerns and the computational requirements of distributed machine learning.

The original PoFL mechanism proposed by Qu et al.~\cite{qu2021proof} represented the first work to employ federated learning as proof of work for blockchain consensus. This groundbreaking approach replaced traditional hash-based PoW with federated learning computation, where energy originally wasted on meaningless PoW puzzles was reinvested into federated learning tasks. The mechanism addressed data privacy leakage through reverse game-based data trading and implemented privacy-preserving model verification using homomorphic encryption and secure two-party computation techniques.

The enhanced protocol, detailed in Algorithm~\ref{alg:enhanced-pofl}, introduces a difficulty adjustment mechanism that adapts mining targets based on model performance metrics rather than traditional hash rates. As shown in line 8, miners with better training performance (lower loss $L_i$) receive easier mining targets through the exponential adjustment factor $e^{-\alpha L_i}$, incentivizing high-quality federated learning contributions.

\begin{algorithm}
\caption{Enhanced Proof of Federated Learning Consensus}
\label{alg:enhanced-pofl}
\begin{algorithmic}[1]
\STATE \textbf{Input:} Previous global model $w^{(t-1)}$, participant local datasets $\mathcal{D}_i$
\STATE \textbf{Mining Phase:}
\FOR{each miner $i$}
   \STATE Train local model: $w_i^{(t)} = \arg\min_{w} \mathcal{L}(\mathcal{D}_i, w^{(t-1)})$
   \STATE Compute training loss: $L_i = \mathcal{L}(\mathcal{D}_i, w_i^{(t)})$
   \STATE Apply differential privacy: $\tilde{w}_i^{(t)} = w_i^{(t)} + \mathcal{N}(0, \sigma^2 I)$
   \STATE Create block with model update and cryptographic proof
   \STATE Attempt to find nonce such that $H(\text{block} || \text{nonce}) < \text{target} \times e^{-\alpha L_i}$
\ENDFOR
\STATE \textbf{Selection:} Choose block with lowest adjusted hash value
\STATE \textbf{Verification:} Network verifies training proofs and model quality
\STATE \textbf{Aggregation:} Winning miner's model becomes basis for next round
\STATE \textbf{Incentive Distribution:} Distribute rewards based on contribution quality
\end{algorithmic}
\end{algorithm}

This performance-based difficulty adjustment considers model metrics including training accuracy, convergence speed, and data quality measures. The dynamic adjustment accounts for miner characteristics such as training sample sizes, non-IID data degrees, and network delays, with Nash-stable convergence mechanisms ensuring optimal federation formation.

This protocol's integration with federated learning aggregation methods presents both unique opportunities and significant challenges. \textsc{FedAvg}~\cite{mcmahan2017communication} compatibility involves synchronizing communication rounds with blockchain block generation, where mining rewards can be tied to contribution to global model improvement. \textsc{FedProx}~\cite{li2020federated} integration can leverage proximal regularization to prevent excessive deviation from the global model, which may be particularly valuable in heterogeneous miner environments. \textsc{FedNova}'s~\cite{wang2020tackling} normalized averaging approach could address varying update frequencies in mining environments, potentially eliminating objective inconsistency problems through variance reduction techniques.

Experimental evaluation demonstrates that PoFL can achieve consensus verification within seconds while maintaining cryptographic security through homomorphic encryption and secure computation protocols. The mechanism successfully incentivizes quality federated learning contributions through reputation-based trading mechanisms~\cite{qu2021proof}. However, like other mining-based approaches, PoFL inherits probabilistic finality characteristics that may limit adoption in enterprise environments requiring deterministic consensus guarantees.

\subsection{Practical Byzantine Fault Tolerance for Federated Learning (FL-PBFT)}

For enterprise deployments requiring deterministic finality and low latency, FL-PBFT adapts classical Byzantine fault tolerance protocols to federated learning requirements. Unlike the probabilistic finality of mining-based approaches, FL-PBFT provides immediate consensus guarantees through committee-based validation, making it particularly suitable for mission-critical applications where model updates must be confirmed with certainty.

FL-PBFT extends traditional PBFT consensus~\cite{castro1999practical} with federated learning-specific validation mechanisms while maintaining the three-phase commit protocol structure. The consensus process operates through specialized phases that integrate model quality assessment with Byzantine fault detection, ensuring both network security and learning efficacy.

The protocol begins with a \textbf{pre-prepare phase} where the designated leader broadcasts model blocks containing aggregated global model parameters. During this phase, participants validate model update authenticity through cryptographic signature verification and perform local training verification to ensure computational work legitimacy. The leader must demonstrate that proposed model updates derive from valid federated learning computations rather than arbitrary parameter modifications.

The \textbf{prepare phase} incorporates federated learning-specific validation mechanisms that distinguish FL-PBFT from traditional Byzantine consensus. Participants perform model quality verification using their local datasets as validation sets, computing performance metrics to assess the proposed global model's effectiveness. The protocol implements honest-training verification that examines training dataset characteristics, local computation times, and gradient magnitudes to detect potential Byzantine behavior. Participants exchange prepare messages only after confirming that the proposed model update meets both cryptographic and learning quality thresholds.

The \textbf{commit phase} confirms network-wide agreement on model blocks while integrating reward distribution mechanisms based on participant contribution quality. Unlike traditional PBFT's uniform treatment of all honest participants, FL-PBFT implements contribution-weighted consensus where participants with higher-quality model updates receive proportionally greater influence in the final decision.

Algorithm~\ref{alg:fl-pbft} presents the complete FL-PBFT consensus protocol, highlighting the integration of Byzantine fault tolerance with federated learning validation.

\begin{algorithm}[!t]
\caption{FL-PBFT Consensus Protocol}
\label{alg:fl-pbft}
\begin{algorithmic}[1]
\STATE \textbf{Input:} Committee nodes $C = \{c_1, c_2, \ldots, c_n\}$, global model $w^{(t-1)}$
\STATE \textbf{Pre-prepare Phase:}
\STATE Leader $c_l$ aggregates local models: $w^{(t)} = \text{Aggregate}(\{w_i^{(t)}\})$
\STATE Leader broadcasts $\langle\text{PRE-PREPARE}, v, n, w^{(t)}, \sigma_l\rangle$
\FOR{each committee node $c_i$}
    \STATE Verify cryptographic signature $\sigma_l$ and model authenticity
    \STATE Validate training computation proofs
\ENDFOR
\STATE \textbf{Prepare Phase:}
\FOR{each committee node $c_i$}
    \STATE Compute validation accuracy: $\text{acc}_i = \text{Evaluate}(w^{(t)}, \mathcal{D}_{val,i})$
    \STATE Perform Byzantine detection: $\text{IsByzantine}(w^{(t)}, \text{gradients})$
    \IF{validation passes}
        \STATE Broadcast $\langle\text{PREPARE}, v, n, \text{acc}_i, \sigma_i\rangle$
    \ENDIF
\ENDFOR
\STATE \textbf{Commit Phase:}
\FOR{each committee node $c_i$}
    \IF{received $2f+1$ valid PREPARE messages}
        \STATE Broadcast $\langle\text{COMMIT}, v, n, \sigma_i\rangle$
    \ENDIF
\ENDFOR
\STATE \textbf{Finality:} Execute model update upon receiving $2f+1$ COMMIT messages
\end{algorithmic}
\end{algorithm}

Byzantine participant detection in FL-PBFT relies on sophisticated gradient analysis methods that examine the statistical properties of model updates. The protocol employs element-wise analysis of gradient vectors to identify model poisoning attacks, computing magnitude distributions and similarity metrics across participant contributions. Advanced detection mechanisms provide adaptive resilience without requiring prior knowledge of malicious participant numbers, enabling real-time anomaly detection during consensus rounds with automated quarantine mechanisms for suspected Byzantine participants.

Similar to the FL-PBFT approach discussed in this section, recent systems have sought to improve the scalability and efficiency of Byzantine fault-tolerant consensus in federated learning. BlockDFL~\cite{qin2024blockdfl} introduces a PBFT-based voting mechanism executed by a small, randomly selected committee of verifiers to reach consensus on global model updates. This process preserves the classical three-phase structure—pre-prepare, prepare, and commit—but restricts participation to a subset of nodes in each round, significantly reducing consensus overhead. By doing so, BlockDFL lowers the communication and computation complexity from $\mathcal{O}(n^2)$ in traditional PBFT to approximately $\mathcal{O}(s)$, where $s$ is the committee size. The BFLC framework~\cite{li2020blockchain} provides an alternative FL-PBFT implementation that separates model update validation from global aggregation through a dual-block architecture. Local model update blocks contain individual participant contributions with cryptographic proofs, while global model blocks store committee-validated aggregations. This separation enables parallel processing of model validation and consensus decisions, further improving system throughput while maintaining Byzantine fault tolerance guarantees.

FL-PBFT's deterministic finality and low latency make it particularly suitable for enterprise federated learning deployments where model consistency and immediate confirmation are critical requirements. However, the protocol's committee-based structure introduces centralization concerns and may limit participation compared to fully decentralized approaches. The trade-off between scalability, security, and decentralization remains an active area of research in Byzantine fault-tolerant federated learning systems.

\subsection{Alternative and Hybrid Consensus Mechanisms}

Beyond the primary consensus mechanisms, the BCFL ecosystem features diverse approaches that address specialized deployment requirements, often combining elements from multiple consensus paradigms to achieve optimal performance for specific use cases.

Contribution-based consensus mechanisms evaluate participants based on their actual contributions to the federated learning process rather than computational work or financial stake. The FedCoin framework~\cite{liu2020fedcoin} exemplifies this approach with its Proof of Shapley (PoSap) protocol, which replaces traditional hash-based mining with Shapley value calculations. Consensus entities compute Shapley values to quantify each participant's marginal contribution to the global model, enabling fair reward distribution based on actual learning value rather than arbitrary computational puzzles.

Building on this foundation, researchers have developed Proof of Collaborative Learning (PoCL)~\cite{sokhankhosh2024proof}, a multi-winner consensus mechanism that redirects blockchain computation directly toward federated learning tasks. Unlike traditional single-winner approaches, PoCL assesses locally trained models and distributes rewards across multiple contributors, improving participation incentives while enhancing system resilience through diversified consensus leadership.

Directed Acyclic Graph (DAG)-based consensus systems replace traditional blockchain structures with DAG architectures to enable parallel processing of consensus decisions. These systems eliminate sequential block ordering constraints, allowing multiple consensus processes to proceed simultaneously. The DAG structure particularly benefits federated learning scenarios with high-frequency model updates, as it reduces consensus latency compared to traditional sequential blockchain approaches.

Reputation-based consensus approaches implement dynamic participant assessment using model quality indicators combined with historical performance factors. Rather than requiring global consensus, these systems employ partial consensus mechanisms where participants maintain local reputation scores for their peers. Consensus decisions incorporate these reputation assessments, creating adaptive systems that respond to participant behavior over time while reducing communication overhead compared to fully synchronized approaches.

\subsection{Consensus Mechanism Selection}

Understanding the trade-offs between consensus mechanisms enables informed architectural decisions for BCFL systems. Each mechanism presents distinct characteristics that align with different deployment requirements and system constraints.

\begin{table}[ht]
\centering
\begin{threeparttable}
\caption{Consensus mechanism comparison for BCFL systems}\label{tab:consensus-comparison}
\begin{tabular}{@{}lccccc@{}}
\toprule
\textbf{Mechanism} & \textbf{Finality} & \textbf{Throughput} & \textbf{Energy} & \textbf{Byzantine} & \textbf{ML Integration} \\
\midrule
Traditional PoW & Probabilistic & Low & Very High & High & None \\
Traditional PoS & Probabilistic & Medium & Low & High & None \\
Proof of Quality & Deterministic & High & Very Low & Medium & Direct \\
Proof of FL & Probabilistic & Medium & Medium & High & Complete \\
FL-PBFT & Deterministic & Very High & Low & High & Integrated \\
\bottomrule
\end{tabular}
\begin{tablenotes}
\item ML Integration: None (no integration), Direct (quality-based), Complete (training-based), Integrated (validation-based)
\end{tablenotes}
\end{threeparttable}
\end{table}

Consensus mechanism selection depends on balancing multiple competing requirements. Latency requirements distinguish real-time applications requiring deterministic finality (FL-PBFT) from applications tolerating probabilistic finality (PoW, PoS, PoFL). Security models determine whether Byzantine fault tolerance is essential for untrusted environments or whether simpler crash fault tolerance suffices for consortium settings. Energy constraints favor computation-efficient mechanisms (PoQ, PBFT variants) over energy-intensive approaches (PoW, PoFL) when consensus is the sole requirement.

Federated learning integration depth represents a critical differentiator. Traditional mechanisms (PoW, PoS) provide blockchain security without ML-specific optimizations, requiring separate model validation layers. Quality-based mechanisms (PoQ) align consensus with model performance but maintain separation between training and consensus phases. Training-integrated approaches (PoFL) can achieve superior energy efficiency when both consensus and model training are needed, by eliminating redundant computation between these processes, though they sacrifice deterministic finality and remain computationally intensive compared to lightweight consensus mechanisms. Committee-based systems offer flexible integration levels through selective ML validation within fast consensus protocols. These fundamental trade-offs guide consensus mechanism selection as the foundational architectural decision that shapes all other BCFL system design choices.

\section{Storage Architecture}

The storage dimension of our taxonomy addresses how BCFL systems manage large machine learning models within blockchain constraints. Traditional blockchain systems impose strict transaction size limits and high storage costs that make direct on-chain storage of neural network parameters challenging. A typical ResNet-50 model with 25 million parameters requires approximately 100MB of storage, while public blockchain networks generally maintain practical transaction size limits in the hundreds of kilobytes.

Storage architectures in BCFL systems typically balance three competing requirements: maintaining cryptographic integrity of model parameters, ensuring high availability for training coordination, and minimizing storage costs across distributed participants.

\subsection{Hybrid Storage Architecture}

\begin{figure}[ht]
    \centering
    \includegraphics[width=1\textwidth]{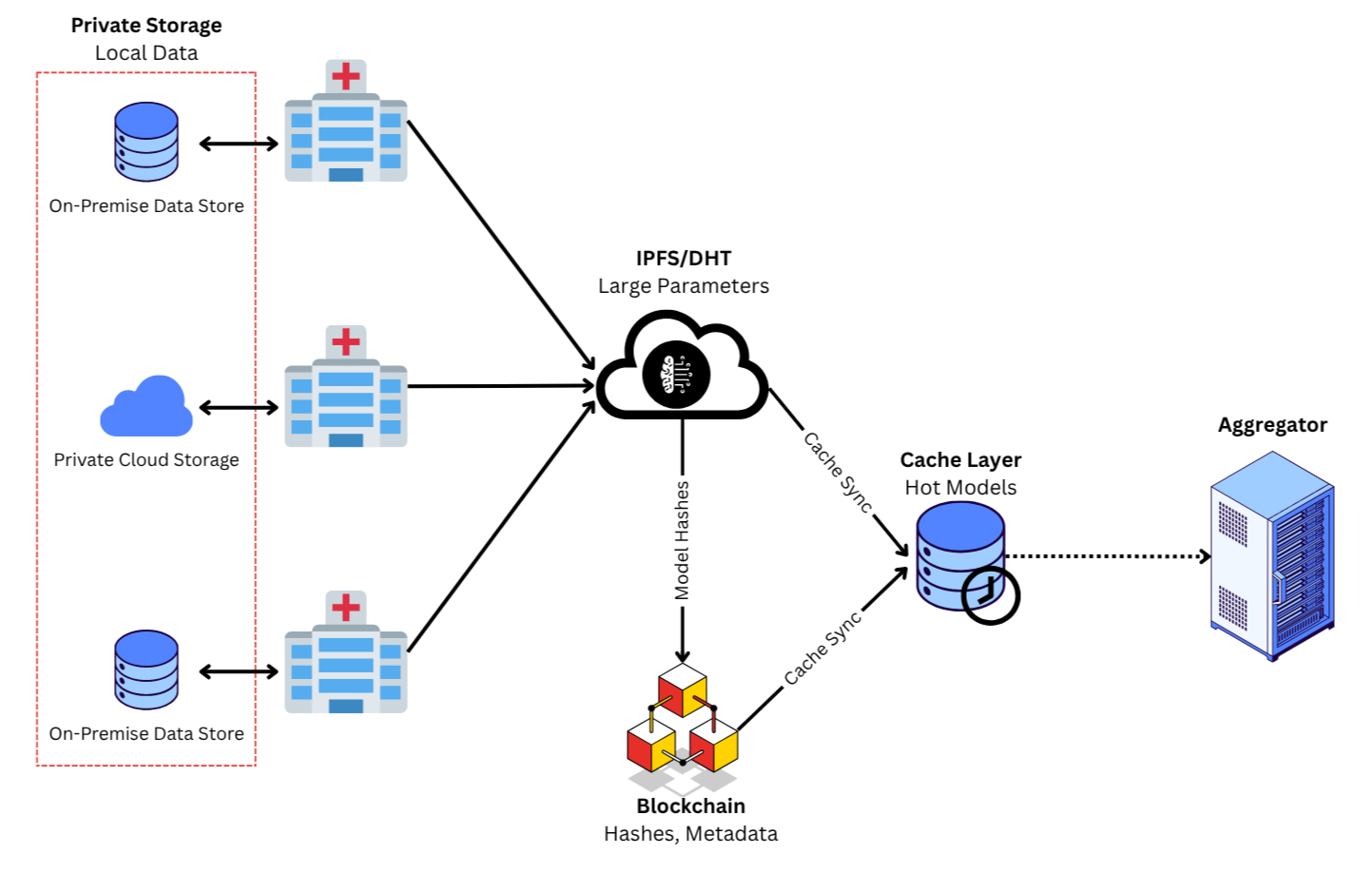}
    \caption{Hybrid storage architecture for BCFL systems showing tiered data management with blockchain integrity, distributed storage scalability, and local caching optimization.}
    \label{fig:storage-architecture}
\end{figure}

Most BCFL implementations adopt tiered storage architectures that strategically distribute data across multiple layers based on size, security requirements, and access patterns. Figure~\ref{fig:storage-architecture} illustrates this multi-tier approach commonly found in BCFL systems.

The blockchain layer serves as the trust anchor, typically storing only critical metadata essential for system integrity. This includes cryptographic hashes of model parameters for tamper detection, smart contract state for access control and participant verification, training round metadata including timestamps and participant contributions, and small model parameters such as bias terms and hyperparameters. Security guarantees provided by on-chain storage extend to other storage tiers through cryptographic linking. By storing model hashes on-chain, participants can verify the integrity of large models stored in distributed systems, creating a chain of trust that preserves blockchain's security properties while enabling scalable storage.

Large model parameters that exceed blockchain capacity are typically stored in distributed hash table (DHT) systems, with IPFS being a common choice that provides content-addressable storage with built-in redundancy~\cite{benet2014ipfs}. The DHT layer handles model weight matrices and large tensors, leveraging content addressing where cryptographic hashes serve as both storage addresses and integrity proofs. Automatic replication across multiple nodes provides fault tolerance against individual node failures, while the peer-to-peer architecture eliminates single points of failure. The distributed nature of DHT storage also provides natural load balancing based on access patterns.

A distributed caching layer optimizes performance for frequently accessed models, including in-memory caches for current global models, content delivery networks for popular model variants, and edge storage for low-latency access. Cache consistency is maintained through invalidation protocols synchronized with blockchain state updates.

Private storage manages participant-specific data through local encrypted datasets, intermediate training results, and participant-specific model variants. End-to-end encryption ensures that sensitive data remains protected while supporting collaborative federated learning aspects. This layer maintains participant sovereignty over sensitive data that participants do not share, even in encrypted form.

\begin{table}[ht]
\centering
\begin{threeparttable}
\caption{Storage tier characteristics and optimization strategies}\label{tab:storage-tiers}
\begin{tabular}{@{}lccccl@{}}
\toprule
\textbf{Tier} & \textbf{Capacity} & \textbf{Latency} & \textbf{Cost} & \textbf{Security} & \textbf{Primary Use Cases} \\
\midrule
On-Chain & $<1$MB & 1-30s & Very High & Maximum & Metadata, hashes, governance \\
DHT/IPFS & Unlimited & 5-60s & Low & High & Model parameters, archives \\
Private Storage & Unlimited & Variable & Medium & Maximum & Local data, gradients \\
Cache Layer & $<100$MB & $<1$s & Medium & Medium & Active models, hot data \\
\bottomrule
\end{tabular}
\begin{tablenotes}
\item Latency represents typical access times under normal network conditions
\end{tablenotes}
\end{threeparttable}
\end{table}

\subsection{Data Integrity and Model Versioning}

Ensuring data integrity across the multi-tier storage architecture presents unique challenges, as model parameters may be distributed across blockchain, DHT, and private storage systems with different security guarantees. BCFL systems address this through layered cryptographic verification protocols that maintain trust relationships between storage tiers while enabling efficient model management. Multi-tier hash verification forms the foundation of cross-layer integrity assurance, where each storage tier maintains cryptographic proofs that can be independently verified against the blockchain's immutable records. When large models are stored in DHT systems, their cryptographic hashes are recorded on-chain, creating a verifiable link between the scalable off-chain storage and the tamper-proof blockchain layer. This approach extends the blockchain's security guarantees to external storage systems without requiring all data to be stored on-chain.

To enable efficient verification and updates of large neural networks, BCFL systems typically organize model parameters into Merkle tree structures. A Merkle tree is a binary tree where each leaf node contains a hash of a data block, and each internal node (intermediate node) contains a hash computed from its child nodes. In the context of neural networks, individual model layers or parameter groups form the leaf nodes, while intermediate nodes represent hierarchical combinations of these components, ultimately producing a single root hash that represents the entire model. This hierarchical structure provides several key advantages for distributed model management: participants can perform partial verification by downloading only specific branches of the tree rather than the entire model, significantly reducing bandwidth requirements for integrity checks; when model updates occur during federated learning rounds, only the affected branches need to be recomputed and propagated, enabling efficient incremental updates; and the tree structure supports parallel verification, as different participants can simultaneously verify different model components.

Beyond ensuring current model integrity, BCFL systems usually maintain comprehensive audit trails that provide complete historical records of model evolution. Each training round creates immutable blockchain records that include the complete participant list with cryptographic identity verification, global model snapshots with their corresponding Merkle root hashes, aggregation quality metrics and validation results, and timestamps linking model versions to specific training epochs. This systematic documentation typically enables several critical capabilities: forensic analysis of model performance degradation, regulatory compliance for auditable AI systems, and research analysis of federated learning convergence patterns. The combination of multi-tier integrity verification, hierarchical model organization, and comprehensive audit trails offers a framework for managing large-scale distributed machine learning models while maintaining the security and transparency benefits of blockchain technology.

\subsection{Performance and Security Considerations}

Storage architecture performance directly impacts federated learning efficiency through model distribution latency and coordination overhead, making optimization strategies critical for practical BCFL deployments. Intelligent caching strategies can leverage federated learning patterns to significantly improve system performance. These approaches typically include predictive model prefetching based on training schedules and participant access patterns, differential update caching that stores only model changes rather than complete parameters, and geographically distributed cache placement to minimize network latency for global participants. Well-designed caching systems can reduce model access latency by substantial margins compared to naive distributed storage approaches, though performance gains vary significantly based on network topology, participant distribution, and model update frequency.

Beyond caching strategies, BCFL systems also need to address the fundamental challenge of efficiently storing and transmitting large neural networks across constrained blockchain and network environments. Model compression techniques such as quantization reduce parameter precision from 32-bit floating point to 8-bit or even binary representations with carefully managed accuracy trade-offs~\cite{jacob2018quantization}, while sparse model encoding exploits the prevalence of near-zero parameters in trained networks to achieve significant storage savings through specialized compression algorithms. These compression methods must balance storage efficiency with computational overhead, as participants need to decompress models for local training while maintaining cryptographic integrity verification across the compressed representations.

Moreover, these performance optimizations need to be implemented within robust security frameworks that enforce participant permissions while supporting collaborative federated learning requirements. Multi-tier access control implements security mechanisms appropriate for each storage layer: smart contract-based blockchain access control manages critical metadata and governance functions, capability-based DHT security enables efficient large-scale model distribution while preventing unauthorized access, and local encryption protects private storage containing sensitive participant data and intermediate training results. The integration of these security layers requires careful coordination to maintain consistent access policies across the distributed system while avoiding performance bottlenecks that could undermine the practical viability of BCFL deployments.

\section{Trust Models}

The trust model dimension of our taxonomy defines participation requirements and access control mechanisms that determine who can join the BCFL network and under what conditions. Trust models establish fundamental security assumptions about participant behavior and shape architectural decisions across all other taxonomic dimensions. Systems with unknown participants require different validation mechanisms than those with pre-verified identities, while the degree of participant vetting determines the feasibility of various consensus and coordination approaches. BCFL systems operate under three primary trust models, each presenting distinct trade-offs between openness, security, and operational complexity. The choice of trust model creates cascading effects throughout the system architecture, from consensus protocol feasibility to storage security requirements and coordination structure viability.

\subsection{Permissionless Trust Model}

Permissionless BCFL systems allow unrestricted participation without prior authorization, mirroring the open access model of public blockchain networks~\cite{nakamoto2008bitcoin,wood2014ethereum}. This approach maximizes decentralization and enables global collaboration, but the absence of participant vetting introduces fundamental security vulnerabilities that significantly complicate system design. The permissionless model operates on three core principles: open participation without gatekeepers, pseudonymous identities secured through cryptographic verification, and distributed governance without centralized approval authorities. While this openness enables unprecedented global collaboration potential, it simultaneously exposes the system to adversarial participants who can exploit the trust-free environment.

The most critical security threats in permissionless environments include Sybil attacks, where malicious actors create multiple fake identities to gain disproportionate influence over consensus decisions, and model poisoning attacks, where adversaries submit carefully crafted corrupted updates designed to degrade global model performance or introduce hidden backdoors. These attack vectors are particularly challenging in federated learning contexts because malicious model updates can be subtle and difficult to detect through traditional cryptographic verification alone. Modern permissionless BCFL systems address these inherent vulnerabilities through multi-layered defense strategies. Economic barriers to participation, such as stake requirements, impose costs on identity creation that make large-scale Sybil attacks financially prohibitive. However, as discussed in Section~\ref{sec:consensus}, traditional Proof of Stake mechanisms prove suboptimal for federated learning applications because financial stake does not correlate with model quality or training data value~\cite{huang2022blockchain}.

Reputation systems provide complementary protection through long-term behavioral tracking, enabling the network to identify and gradually marginalize participants who consistently contribute low-quality or malicious updates. Zero-knowledge proof integration offers privacy-preserving verification capabilities, allowing validation of computational correctness and model quality without exposing sensitive parameters or training data. The zkFL framework~\cite{xing2023zero} demonstrates this approach, enabling blockchain miners to verify federated learning computations while maintaining complete parameter privacy. 

Real-world deployments have validated the practical viability of permissionless approaches despite their security complexities. FLock.io~\cite{flock2024mainnet} represents the most mature deployment, operating on Base (Ethereum Layer 2) with over 700,000 users and implementing a Proof-of-Federated-Work consensus that validates model updates through cryptographic proofs and distributed verification across validator nodes. Similarly, the Artificial Superintelligence Alliance~\cite{asi2024intelligence} demonstrates multi-chain federated learning across Ethereum and Arbitrum, utilizing reputation-weighted consensus where node trustworthiness is computed from historical model quality metrics and cryptographic attestations. Despite these promising implementations, the consensus mechanisms suitable for permissionless environments remain fundamentally constrained by the trustless participation model. Traditional Proof of Work approaches face the energy consumption and latency challenges discussed in Section~\ref{sec:consensus}, while quality-based mechanisms like Proof of Quality require trusted validation datasets that conflict with the trustless assumptions of permissionless systems. This tension between openness and security verification represents an ongoing research challenge in permissionless BCFL design.

\subsection{Consortium Trust Model}

Consortium models restrict participation to pre-approved organizations that form collaborative alliances, balancing the openness benefits of permissionless systems with the control and efficiency advantages of restricted access. This intermediate approach proves particularly suitable for competitive industries where organizations recognize mutual benefits from collaboration while maintaining strict control over participation. Consortium implementations feature sophisticated governance structures through several key mechanisms: vetted participants undergo rigorous identity verification processes that establish organizational legitimacy and technical capability; multi-signature smart contracts enforce approval workflows that distribute control among consortium members, preventing unilateral decisions while maintaining operational efficiency; and weighted voting systems allocate influence based on quantifiable contributions such as data volume, computational resources, or historical participation quality.

The technical implementation of consortium trust typically involves multi-signature participant onboarding requiring approval from multiple existing members, consortium-specific smart contracts managing voting and resource allocation, and optimized consensus mechanisms tailored to the known participant set. Consortium models enable the deployment of consensus mechanisms that would be impractical in permissionless environments. The known participant set allows for efficient Byzantine fault-tolerant protocols like FL-PBFT, as discussed in Section~\ref{sec:consensus}, while the established trust relationships support quality-based consensus mechanisms without requiring trustless validation infrastructure.

The pharmaceutical industry provides the most prominent example of successful consortium implementation through MELLODDY, which unites 10 major companies including Amgen, AstraZeneca, and Novartis. This consortium collaboratively trains models on over 2.6 billion experimental data points while maintaining strict intellectual property protection, achieving 4-20\% improvement in predictive performance compared to individual company efforts~\cite{heyndrickx2023melloddy}. The consortium model's success in this highly competitive industry demonstrates its effectiveness for scenarios requiring both collaboration and strict participation control. Consortium trust models also support hierarchical coordination structures more effectively than permissionless systems. The established organizational relationships provide natural boundaries for hierarchical grouping, while the known participant characteristics enable optimized coordination protocols.

\subsection{Permissioned Trust Model}

Fully permissioned systems require explicit authorization for all participants, typically managed by a central authority or governance board. This model provides maximum control and regulatory compliance capabilities while sacrificing the decentralization benefits that motivate blockchain adoption, making it most suitable for enterprise deployments and highly regulated industries. Permissioned architectures implement comprehensive identity and access management through several integrated mechanisms: Know Your Customer (KYC) procedures establish participant identity and compliance status through traditional verification processes; role-based access control (RBAC) provides fine-grained permission management that can differentiate between data providers, model trainers, aggregators, and validators; certificate-based authentication leverages Public Key Infrastructure (PKI) to establish cryptographic identity verification; and centralized participant management provides comprehensive oversight capabilities including real-time monitoring and immediate revocation authority.

The controlled participant environment in permissioned systems enables significant performance optimizations. Trusted participant assumptions allow for streamlined consensus mechanisms that eliminate Byzantine fault tolerance overhead in favor of simpler crash fault tolerance protocols. Communication costs decrease substantially due to the elimination of adversarial behavior verification, while data poisoning attack detection improves through comprehensive participant monitoring and accountability mechanisms. Enterprise implementations demonstrate sophisticated integration with existing cloud and enterprise infrastructure, with financial services leading adoption in this space. JPMorgan's FedSyn framework combines the bank's Quorum blockchain platform with federated learning to enable multiple financial institutions to generate synthetic data for AI model training while preserving customer privacy through differential privacy and secure aggregation~\cite{behera2022fedsyn}.

However, the enhanced security and performance of permissioned systems comes at the cost of increased centralization and potential single points of failure. The central authority responsible for participant management represents a vulnerability that contradicts blockchain's fundamental decentralization principles. Enterprise deployments must carefully balance these trade-offs based on specific regulatory requirements and organizational constraints. Permissioned models prove particularly valuable for applications requiring regulatory compliance in sectors such as healthcare, finance, and critical infrastructure. NATO's evaluation of blockchain-based federated learning for secure military communications exemplifies this application domain, where strict access control outweighs decentralization concerns~\cite{wrona2018does}.

\subsection{Trust Model Selection and Hybrid Approaches}

\begin{table}[ht]
\centering
\begin{threeparttable}
\caption{Trust model comparison across key characteristics}\label{tab:trust-comparison}
\begin{tabular}{@{}lccc@{}}
\toprule
\textbf{Characteristic} & \textbf{Permissionless} & \textbf{Consortium} & \textbf{Permissioned} \\
\midrule
Entry Barriers & None & Organizational Vetting & Strict Authorization \\
Identity Verification & Pseudonymous & Known Entities & Full KYC/Compliance \\
Consensus Efficiency & Low (Byzantine) & Medium (Optimized) & High (Trusted) \\
Regulatory Compliance & Challenging & Moderate & Comprehensive \\
Decentralization Level & Maximum & Balanced & Minimal \\
Blockchain Performance (TPS) & 10-20 & 200-1000 & 1000+ \\
Energy Efficiency & Low (Mining) & Medium (BFT) & High (Optimized) \\
Sybil Resistance & Cryptoeconomic & Admission Control & Identity Verification \\
Suitable Consensus & PoW, PoFL, PoQ* & FL-PBFT, PoQ & Optimized PBFT \\
\bottomrule
\end{tabular}
\begin{tablenotes}
\item TPS = Transactions Per Second; PoW = Proof of Work; PoFL = Proof of Federated Learning
\item PoQ* = Proof of Quality with trustless validation; BFT = Byzantine Fault Tolerant
\end{tablenotes}
\end{threeparttable}
\end{table}

Comparative analysis reveals that no single trust model optimally addresses all BCFL requirements, driving innovation in hybrid approaches that combine strengths from multiple models while mitigating their individual weaknesses. These hybrid architectures recognize that different aspects of a BCFL system may benefit from different trust assumptions.

Multi-tier trust architectures implement varying trust requirements for different network roles within the same system. The AgriFLChain framework implements this approach with three distinct layers: an open contribution tier allowing broad data provider participation, a vetted aggregator layer with reputation-based selection mechanisms, and a governance layer employing multi-signature smart contracts for critical decisions~\cite{agriflchain2025}. This stratified approach enables broad participation while maintaining control over critical system functions.

Dynamic trust models implement context-aware adaptation that can modify trust requirements based on network conditions, participant behavior, or application requirements. These systems can transition between different consensus mechanisms or participation models as circumstances change, providing flexibility that static trust models cannot achieve.

Cross-chain implementations enable organizations to participate in multiple blockchain networks simultaneously, leveraging different trust models for different aspects of the federated learning process. This approach allows organizations to maintain permissioned coordination for sensitive operations while participating in consortium or permissionless networks for broader collaboration.

The trust model selection fundamentally shapes BCFL system capabilities and determines the feasibility of specific consensus mechanisms, coordination structures, and storage architectures. As summarized in Table \ref{tab:trust-comparison}, each trust model presents distinct trade-offs across entry barriers, consensus efficiency, and regulatory compliance requirements. The continued development of hybrid approaches suggests that future BCFL systems will increasingly support dynamic trust model adaptation based on changing operational requirements, regulatory environments, and participant ecosystems.

\section{A Case Study: Hands-On Federated Learning with \textsc{TrustMesh}}

To illustrate the practical deployment and operation of blockchain-enabled federated learning systems, we present a hands-on case study using \textsc{TrustMesh}~\cite{rangwala2025trustmesh}, a blockchain-based distributed computing framework\footnote{Source code available at \url{https://github.com/Cloudslab/TrustMesh-FL}}. This case study demonstrates the complete process of setting up and running a federated learning task in a blockchain environment, showing how the coordination differs from traditional federated learning approaches and what benefits this provides in practice.

\subsection{Understanding the Edge Federated Learning Setup}

Our demonstration uses an edge federated learning scenario where IoT devices across different locations want to collaboratively train a machine learning model without sharing their private data. This represents a common real-world situation involving smart cameras in different buildings, industrial sensors across multiple facilities, or mobile devices in various geographic regions that each have local data but would benefit from a shared model.

In traditional edge federated learning, these distributed devices would need to trust a central coordinator managed by a single organization. With blockchain-enabled federated learning, the coordination is handled by a decentralized network of edge compute nodes where no single party controls the process. The system architecture comprises three main components:
\begin{itemize}
    \item \textbf{IoT Edge Devices}: Low power resource-constrained devices representing sensors, cameras, or mobile devices that hold private local data
    \item \textbf{Edge Compute Nodes}: A network of more powerful edge servers that perform training computations and validate blockchain transactions
    \item \textbf{Blockchain Coordination}: The same edge compute nodes maintain a shared ledger that coordinates the entire federated learning process transparently
\end{itemize}

In this architecture, the edge compute nodes serve dual roles—they perform machine learning computations as well as maintain blockchain consensus. This design eliminates the need for a separate trusted coordinator while ensuring all coordination decisions are transparent and verifiable by all participating edge devices.

\subsection{The Learning Task: MNIST Digit Classification}

For this demonstration, we use the classic MNIST handwritten digit recognition dataset distributed across edge IoT devices. Each device holds a different subset of the data, simulating a scenario where different edge locations have specialized but complementary datasets. The dataset is partitioned such that each of the five IoT devices in our testbed contains samples from only two digit classes (approximately 10,000-12,000 images per device), creating a highly non-IID distribution.

This data distribution is intentionally uneven—each edge device can only recognize certain digits well when training alone. To build a model that recognizes all digits accurately, the devices must collaborate while keeping their data private. This scenario closely mirrors real-world edge computing situations where different locations such as retail stores, manufacturing plants, or geographic regions possess complementary datasets that would benefit from collaborative learning.

\subsection{How Blockchain Coordination Works}

\begin{figure}[!ht]
    \centering
    \includegraphics[width=1\textwidth]{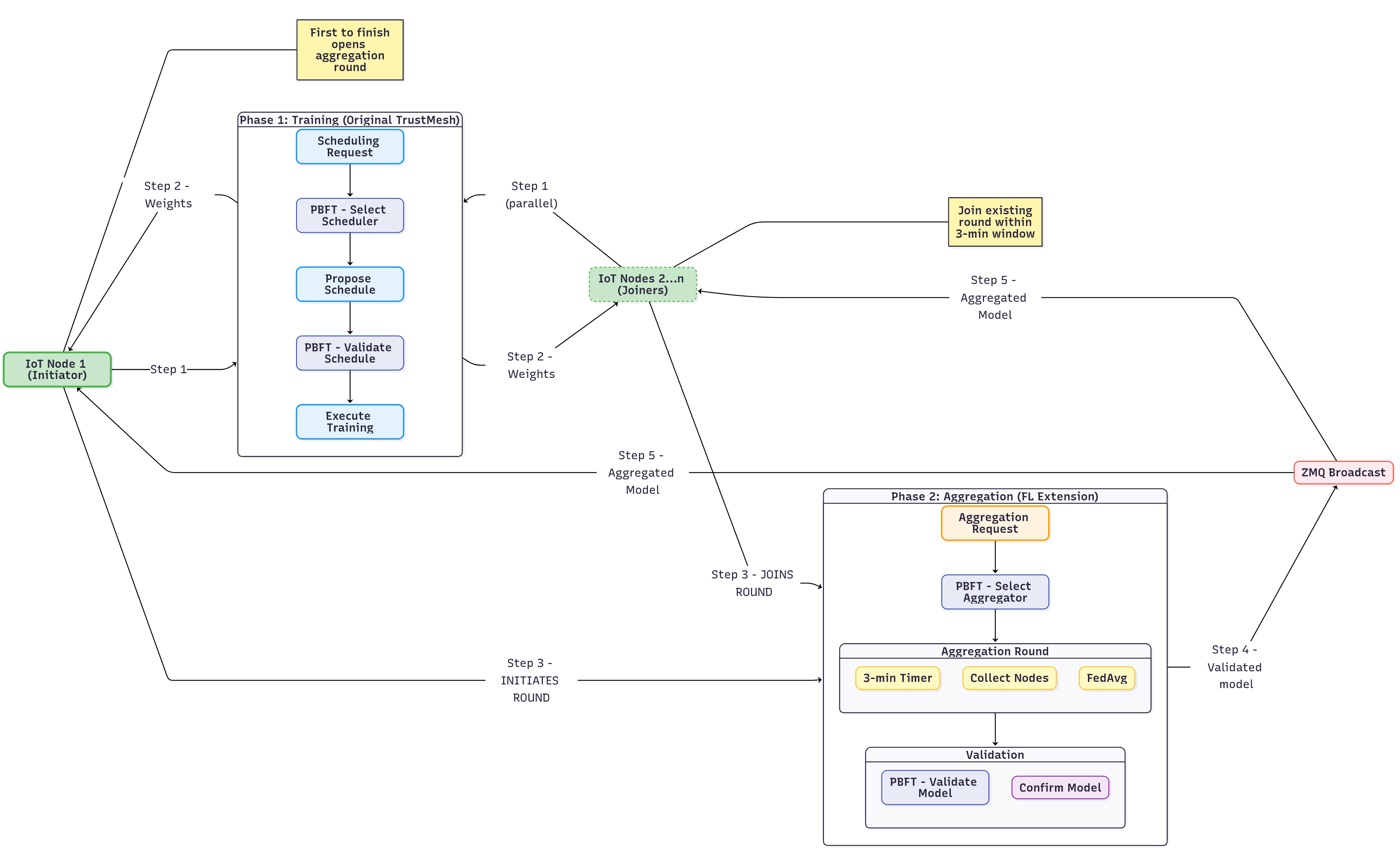}
    \caption{\textsc{TrustMesh} Federated Learning Architecture}
    \label{fig:trustmesh}
\end{figure}

The blockchain coordination process replaces the central server with a transparent, automated system that operates through two distinct phases. As depicted in Figure~\ref{fig:trustmesh}, the federated learning execution in \textsc{TrustMesh} follows a distinctive coordination pattern that leverages blockchain technology for verification and audit trail maintenance.

\paragraph{Training Round Coordination}
When an edge IoT device wants to start a new learning round, it sends a request to the blockchain network. The system then follows \textsc{TrustMesh}'s multi-phase consensus protocol to coordinate the training. First, the blockchain network uses a deterministic resource-based method to assign a scheduler node through consensus—all edge compute nodes examine the available resources and agree on which node should act as the scheduler. The designated scheduler then proposes a specific task execution schedule that details how the training should be performed. This proposed schedule is validated by all blockchain nodes through consensus before training is actually triggered. Once the schedule is approved, training begins according to the agreed-upon plan, and all training assignments and decisions are recorded on the blockchain for complete transparency.

\paragraph{Model Aggregation Coordination}
After individual training is complete, the nodes coordinate to combine their models. IoT devices submit their trained model updates to the blockchain, and the system opens a three-minute collection window to gather contributions from all participants. Another automated selection process chooses which edge compute node will perform the model averaging. For this case study, we use the \textsc{FedAvg} algorithm~\cite{mcmahan2017communication}, though \textsc{TrustMesh} allows users to implement custom aggregation algorithms based on their specific requirements. The aggregator combines all submitted models, and the final global model is validated by the network and committed to the system before being broadcast to all participating IoT devices for the next round.

It is important to note that \textsc{TrustMesh} provides flexibility in how IoT devices participate in the federated learning process. While the system can deploy training tasks to edge compute nodes on behalf of resource-constrained IoT devices, devices with sufficient computational capabilities can choose to perform local training independently and participate only in the aggregation phase. This design accommodates varying IoT device capabilities while maintaining the federated learning paradigm.

\subsection{Running the System: A Step-by-Step Walkthrough}

\subsubsection{Initial Setup}

The first step involves deploying the federated learning application across the edge network. This process utilizes \textsc{TrustMesh}'s blockchain-based application management to ensure all nodes receive identical application configurations.

\begin{lstlisting}[language=bash, caption=Application Deployment, keywordstyle=\ttfamily]
# Deploy the federated learning Docker image
$ kubectl exec -it network-management-console-xxx -c application-deployment-client -- bash
$ python docker_image_client.py deploy_image \
   federated-training-task.tar.gz app_requirements.json
   
INFO: Processing Docker image from tar: federated-training-task.tar.gz
INFO: Image push completed successfully
INFO: Transaction created and submitted to blockchain
Deployment submitted. Application ID: a3aaf5d5-460d-4ee5-a427-e057d3df4e97
Status: SUCCESS
\end{lstlisting}

Next, a workflow must be created that defines how the federated learning process will be coordinated:

\begin{lstlisting}[language=bash, caption=Workflow Creation, keywordstyle=\ttfamily]
# Create the federated learning workflow
$ kubectl exec -it network-management-console-xxx -c workflow-creation-client -- bash
$ python workflow_creation_client.py dependency_graph.json

INFO: Creating new workflow
Workflow ID: c954f15f-6df3-4ff2-8e33-c77c29800006
Status: Transaction submitted successfully
\end{lstlisting}

\subsubsection{Executing the Learning Process}

Each IoT device joins the federated learning process by executing the simulation script with the generated workflow ID. The system automatically detects each device's identity and assigns appropriate data partitions:

\begin{lstlisting}[language=bash, caption=Starting IoT Node Participation, keywordstyle=\ttfamily]
# Start federated learning on each IoT device
$ kubectl exec -it iot-0-xxx -c iot-node -- bash
$ python mnist-federated-learning-simulation.py \
    --workflow-id c954f15f-6df3-4ff2-8e33-c77c29800006 \
    --max-rounds 10

MNIST FEDERATED LEARNING NODE
Node ID: iot-0
Workflow ID: c954f15f-6df3-4ff2-8e33-c77c29800006
- Assigned Classes: [0, 1]
- Training Samples: 10132
- Test Samples: 2533
\end{lstlisting}

\subsubsection{Observing the Two-Phase Coordination}

Once started, the system demonstrates its distinctive two-phase coordination process. Each learning round follows a predictable pattern that showcases the blockchain's role in maintaining transparency and consensus.

\paragraph{Phase 1: Training Coordination}
The training phase begins when an IoT device prepares its local data slice and submits it to the blockchain network for processing:

\begin{lstlisting}[language=bash, caption=Training Phase Execution, basicstyle=\footnotesize]
PHASE 1: TRAINING PHASE
DATA PREPARATION: Preparing training data for round 1
- Data slice: 500 samples
- Assigned classes: [0, 1]
- Class distribution: {0: 242, 1: 258}

BLOCKCHAIN TRANSACTION: Submitting training data to TrustMesh
Generated schedule ID: 28e090f9-2525-4e3d-a760-9672bd64d2bc
Batch submitted successfully

TRAINING COMPLETED SUCCESSFULLY
- Wait duration: 19.1s
- Received weights: 10 layers
- Training completed by edge compute node
\end{lstlisting}

The consistent 18-22 second training completion times demonstrate the system's predictable performance characteristics. Each training request generates a unique schedule ID that enables complete traceability of computational work assignment and execution across the distributed edge network.

\paragraph{Phase 2: Aggregation Coordination}
The aggregation phase coordinates multiple trained models into a single global model through blockchain consensus:

\begin{lstlisting}[language=bash, caption=Aggregation Phase Execution, basicstyle=\footnotesize]
PHASE 2: AGGREGATION PHASE
BLOCKCHAIN TRANSACTION: Submitting aggregation request
- Weight layers: 10
- Node classes: [0, 1]
- Training samples used: 10132

AGGREGATED MODEL RECEIVED
- Round number: 2
- Aggregator node: sawtooth-compute-node-2
- Participating nodes: ['iot-2', 'iot-0']
- Blockchain validation score: 0.8006
- Aggregation wait duration: 71.1s

AGGREGATED MODEL RECEIVED
- Round number: 3
- Aggregator node: sawtooth-compute-node-3
- Participating nodes: ['iot-1', 'iot-0', 'iot-2']
- Blockchain validation score: 0.8447
\end{lstlisting}

The aggregation phase demonstrates several key blockchain coordination benefits. The system automatically selects aggregator nodes (sawtooth-compute-node-2, sawtooth-compute-node-3) through consensus rather than predetermined assignment. The three-minute collection window accommodates asynchronous participation, allowing nodes to contribute even with varying connectivity patterns. Most importantly, blockchain validation scores provide consensus-based quality metrics that all participants can verify.

\subsubsection{Multi-Node Collaboration Patterns}

As additional IoT devices join the collaborative learning process, the blockchain coordination seamlessly scales to accommodate increased participation. When iot-1 (handling digits 2,3) and iot-2 (handling digits 4,5) join the workflow, they follow identical coordination processes while contributing their specialized knowledge:

\begin{lstlisting}[language=bash, caption=Multi-Node Coordination Example, basicstyle=\footnotesize]
# iot-1 joins with different digit classes
Node ID: iot-1
- Assigned Classes: [2, 3]
- Training Samples: 9671

# iot-2 contributes additional diversity
Node ID: iot-2
- Assigned Classes: [4, 5]
- Training Samples: 9010

AGGREGATED MODEL RECEIVED
- Participating nodes: ['iot-0', 'iot-1', 'iot-2']
- Aggregator node: sawtooth-compute-node-3
- Blockchain validation score: 0.8447
\end{lstlisting}

This demonstrates how the blockchain coordination enables asynchronous participation—nodes can join at different times and the system automatically coordinates their participation in ongoing rounds. The increasing validation scores (0.7778 → 0.8006 → 0.8447) reflect improving model quality as more diverse data sources contribute to the collaborative learning process.

\subsection{Performance Analysis and Key Insights}

The practical deployment reveals compelling evidence of successful collaborative learning despite challenging non-IID data distribution. Individual nodes possess specialized datasets containing only their assigned digit classes, yet through blockchain-coordinated collaboration achieve improved recognition performance on their local test sets while maintaining complete data privacy.

\subsubsection{Learning Effectiveness Under Non-IID Conditions}

The learning progression demonstrates effective knowledge transfer across the federated network. Node \texttt{iot-0} (digits 0,1) progressed from 0\% baseline to 96.76\% accuracy by round 3, while \texttt{iot-2} (digits 4,5) achieved 89.39\% accuracy with balanced performance across both assigned classes (93.06\% for class 4, 85.55\% for class 5). Node \texttt{iot-1} (digits 2,3) initially reached 50.29\% with perfect class 2 recognition but required additional rounds for class 3 improvement. This represents successful collaborative learning despite the highly non-IID data distribution that typically challenges federated learning systems.

The consensus-based validation scores (0.7778 to 0.8006 to 0.8447) are computed by the blockchain network using a global validation dataset, providing system-wide quality assessment that all participants can independently verify. Meanwhile, individual nodes evaluate model performance on their own local test data, showing how well the global model performs on each node's specialized digit classes.

\subsubsection{Distinctive Operational Characteristics}

The blockchain coordination demonstrates three key advantages over traditional centralized approaches. 

\begin{enumerate}
    \item \textbf{Complete transparency} enables participants to verify every coordination decision through automatically generated schedule IDs and transaction logs. The system demonstrates automatic assignment of different compute nodes (sawtooth-compute-node-2, sawtooth-compute-node-3) based on resource availability and consensus rather than predetermined hierarchies.
    \item \textbf{Asynchronous resilience} accommodates varying participation patterns critical for distributed IoT environments. Nodes can join at different times and participate asynchronously through the three-minute collection window, while random inter-round pauses (148.1s, 152.6s, 231.9s) simulate realistic IoT availability patterns without disrupting learning progress.
    \item \textbf{Predictable performance overhead} enable reliable deployment planning. Training coordination consistently requires 18-22 seconds, while aggregation coordination varies from 53-200 seconds based on the time-based collection window design. Complete federated learning rounds require 75-220 seconds. The time-based aggregation approach prioritizes fault tolerance over speed—using fixed collection windows to accommodate node dropouts and new participants joining mid-process, rather than waiting for all initially registered nodes to respond. This design choice trades some latency for enhanced resilience in dynamic edge environments where node availability cannot be guaranteed.

\end{enumerate}

\subsubsection{Practical Applicability}

This deployment experience reveals three primary use cases where blockchain-enabled edge federated learning provides significant advantages: trustless multi-party collaborations where participants lack established trust relationships, compliance-driven environments mandating comprehensive audit trails, and extended collaborative deployments where the initial coordination complexity is offset by long-term operational benefits. The system's time-based aggregation and consensus mechanisms introduce coordination latency but provide unique capabilities unavailable in centralized approaches—particularly fault tolerance for dynamic node participation, complete transparency of all coordination decisions, and elimination of single points of failure. For scenarios requiring these guarantees—such as industrial IoT deployments, multi-organizational collaborations, and regulatory compliance environments—this approach provides a practical path forward for secure, auditable, and decentralized machine learning. This case study demonstrates that blockchain-enabled federated learning can achieve collaborative learning goals while providing the additional trust and resilience guarantees needed for trustless, multi-party scenarios where traditional centralized coordination would be infeasible or unacceptable.

\section{Challenges and Future Research Directions}

While BCFL systems have demonstrated practical viability across numerous domains, several fundamental challenges remain that limit their broader adoption and scalability potential. The most pressing limitations emerge from the intersection of blockchain consensus requirements and federated learning coordination demands. Current consensus mechanisms face inherent trade-offs between decentralization, security, and throughput that become particularly pronounced in large-scale deployments. As participant numbers grow, the communication overhead and validation complexity of maintaining Byzantine fault tolerance create bottlenecks that constrain system scalability. These challenges are compounded by storage architecture limitations, where the tension between blockchain's transaction size constraints and neural networks' large parameter spaces requires increasingly complex hybrid solutions that introduce their own performance and security considerations.

Beyond scalability concerns, the unique threat model of BCFL systems creates novel security vulnerabilities that traditional blockchain or federated learning systems do not face individually. The combination of public blockchain metadata with private federated learning updates creates new attack vectors for inference and model poisoning that exploit the intersection of these technologies. The immutable nature of blockchain records, while providing valuable audit trails, can also perpetuate the effects of successful attacks across the entire network. Furthermore, the emerging quantum computing threat poses existential risks to the cryptographic foundations underlying both blockchain consensus and secure aggregation protocols.

The economic dynamics of collaborative learning present equally complex challenges for sustainable BCFL deployment. Current incentive mechanisms struggle to fairly assess participant contributions in federated learning environments characterized by non-IID data distributions and varying computational capabilities. The operational costs associated with blockchain consensus and verification must be balanced against the benefits of collaborative learning, creating economic barriers that particularly affect resource-constrained participants who might otherwise benefit most from collaborative approaches.

Addressing these multifaceted challenges requires coordinated research across several critical directions. Consensus mechanism innovation must focus on protocols that maintain security guarantees while achieving better scalability characteristics, potentially through hierarchical approaches or novel committee-based designs. Storage solutions need further integration of compression techniques with cryptographic verification to enable efficient management of large models. Security research must prioritize post-quantum cryptographic integration and privacy-preserving validation mechanisms that can detect sophisticated attacks in real-time. Finally, mechanism design research should develop theoretically grounded approaches to contribution assessment and incentive alignment that ensure fair reward distribution while maintaining economic sustainability for diverse participant ecosystems.

\section{Summary}

This chapter presented a comprehensive analysis of blockchain-enabled federated learning systems through a systematic four-dimensional taxonomy encompassing coordination structures, consensus mechanisms, storage architectures, and trust models. We demonstrated how architectural choices across these dimensions serve specific deployment requirements, from consortium healthcare networks requiring comprehensive audit trails to fully decentralized systems demanding maximum censorship resistance. The evolution from traditional blockchain consensus to specialized federated learning protocols—including Proof of Quality and Proof of Federated Learning—illustrates how computational work can be repurposed from arbitrary cryptographic puzzles to productive machine learning tasks.

The detailed \textsc{TrustMesh} case study validated practical BCFL deployment feasibility, demonstrating effective collaborative learning across IoT devices with highly non-IID data distributions while maintaining complete transparency and eliminating single points of failure. Real-world deployments across healthcare, finance, and IoT domains have proven that BCFL systems can achieve comparable accuracy to centralized approaches while providing enhanced security guarantees and enabling new models of collaborative intelligence. Despite ongoing challenges in scalability, security, and regulatory compliance, BCFL represents a fundamental advancement in trustless multi-party machine learning that unlocks collaborative intelligence opportunities previously impossible due to trust, privacy, or coordination constraints.

\bibliographystyle{IEEEtran}
\bibliography{bcfl_references}

\begin{thebibliography}{10}
\providecommand{\url}[1]{#1}
\csname url@samestyle\endcsname
\providecommand{\newblock}{\relax}
\providecommand{\bibinfo}[2]{#2}
\providecommand{\BIBentrySTDinterwordspacing}{\spaceskip=0pt\relax}
\providecommand{\BIBentryALTinterwordstretchfactor}{4}
\providecommand{\BIBentryALTinterwordspacing}{\spaceskip=\fontdimen2\font plus
\BIBentryALTinterwordstretchfactor\fontdimen3\font minus \fontdimen4\font\relax}
\providecommand{\BIBforeignlanguage}[2]{{%
\expandafter\ifx\csname l@#1\endcsname\relax
\typeout{** WARNING: IEEEtran.bst: No hyphenation pattern has been}%
\typeout{** loaded for the language `#1'. Using the pattern for}%
\typeout{** the default language instead.}%
\else
\language=\csname l@#1\endcsname
\fi
#2}}
\providecommand{\BIBdecl}{\relax}
\BIBdecl

\bibitem{mcmahan2017communication}
B.~McMahan, E.~Moore, D.~Ramage, S.~Hampson, and B.~A. y~Arcas, ``Communication-efficient learning of deep networks from decentralized data,'' in \emph{Artificial intelligence and statistics}.\hskip 1em plus 0.5em minus 0.4em\relax PMLR, 2017, pp. 1273--1282.

\bibitem{tariq2023patient}
\BIBentryALTinterwordspacing
R.~Tariq and P.~Hackert, \emph{Patient Confidentiality}.\hskip 1em plus 0.5em minus 0.4em\relax Treasure Island (FL): StatPearls Publishing, 2023. [Online]. Available: \url{https://www.ncbi.nlm.nih.gov/books/NBK519540}
\BIBentrySTDinterwordspacing

\bibitem{kairouz2021advances}
P.~Kairouz \emph{et~al.}, ``Advances and open problems in federated learning,'' \emph{Foundations and trends in machine learning}, vol.~14, no. 1--2, pp. 1--210, 2021.

\bibitem{mothukuri2021survey}
V.~Mothukuri, R.~M. Parizi, S.~Pouriyeh, Y.~Huang, A.~Dehghantanha, and G.~Srivastava, ``A survey on security and privacy of federated learning,'' \emph{Future Generation Computer Systems}, vol. 115, pp. 619--640, 2021.

\bibitem{li2022blockchain}
D.~Li \emph{et~al.}, ``Blockchain for federated learning toward secure distributed machine learning systems: a systemic survey,'' \emph{Soft Computing}, vol.~26, no.~9, pp. 4423--4440, 2022.

\bibitem{nguyen2021federated}
D.~C. Nguyen \emph{et~al.}, ``Federated learning meets blockchain in edge computing: Opportunities and challenges,'' \emph{IEEE Internet of Things Journal}, vol.~8, no.~16, pp. 12\,806--12\,825, 2021.

\bibitem{castro1999practical}
M.~Castro and B.~Liskov, ``Practical byzantine fault tolerance,'' in \emph{OsDI}, vol.~99, 1999, pp. 173--186.

\bibitem{kang2019incentive}
J.~Kang, Z.~Xiong, D.~Niyato, S.~Xie, and J.~Zhang, ``Incentive mechanism for reliable federated learning: A joint optimization approach to combining reputation and contract theory,'' \emph{IEEE Internet of Things Journal}, vol.~6, pp. 10\,700--10\,714, 2019.

\bibitem{rangwala2025trustmesh}
M.~Rangwala and R.~Buyya, ``Trustmesh: A blockchain-enabled trusted distributed computing framework for open heterogeneous iot environments,'' in \emph{IEEE 22nd International Conference on Software Architecture (ICSA)}, 2025, pp. 131--141.

\bibitem{ying2024bitfl}
C.~Ying \emph{et~al.}, ``Bit-fl: Blockchain-enabled incentivized and secure federated learning framework,'' \emph{IEEE Transactions on Mobile Computing}, 2024.

\bibitem{majeed2019flchain}
U.~Majeed and C.~S. Hong, ``Flchain: Federated learning via mec-enabled blockchain network,'' in \emph{20th Asia-Pacific Network Operations and Management Symposium (APNOMS)}.\hskip 1em plus 0.5em minus 0.4em\relax IEEE, 2019, pp. 1--4.

\bibitem{chai2020hierarchical}
H.~Chai, S.~Leng, Y.~Chen, and K.~Zhang, ``A hierarchical blockchain-enabled federated learning algorithm for knowledge sharing in internet of vehicles,'' \emph{IEEE Transactions on Intelligent Transportation Systems}, vol.~22, no.~7, pp. 3975--3986, 2020.

\bibitem{sarhan2022hbfl}
M.~Sarhan, W.~W. Lo, S.~Layeghy, and M.~Portmann, ``Hbfl: A hierarchical blockchain-based federated learning framework for collaborative iot intrusion detection,'' \emph{Computers and Electrical Engineering}, vol. 103, p. 108379, 2022.

\bibitem{shayan2020biscotti}
M.~Shayan, C.~Fung, C.~J. Yoon, and I.~Beschastnikh, ``Biscotti: A blockchain system for private and secure federated learning,'' \emph{IEEE Transactions on Parallel and Distributed Systems}, vol.~32, no.~7, pp. 1513--1525, 2020.

\bibitem{qin2024blockdfl}
Z.~Qin, X.~Yan, M.~Zhou, and S.~Deng, ``Blockdfl: A blockchain-based fully decentralized peer-to-peer federated learning framework,'' in \emph{Proceedings of the ACM Web Conference}, 2024, pp. 2914--2925.

\bibitem{nakamoto2008bitcoin}
S.~Nakamoto, ``Bitcoin: A peer-to-peer electronic cash system,'' \url{https://bitcoin.org/bitcoin.pdf}, 2008.

\bibitem{wood2014ethereum}
G.~Wood, ``Ethereum: A secure decentralised generalised transaction ledger,'' \emph{Ethereum project yellow paper}, vol. 151, pp. 1--32, 2014.

\bibitem{huang2022blockchain}
J.~Huang, L.~Kong, G.~Chen, Q.~Xiang, X.~Chen, and X.~Liu, ``Blockchain-based federated learning: A systematic survey,'' \emph{IEEE Network}, vol.~37, no.~6, pp. 150--157, 2022.

\bibitem{chen2021robust}
H.~Chen, S.~A. Asif, J.~Park, C.-C. Shen, and M.~Bennis, ``Robust blockchained federated learning with model validation and proof-of-stake inspired consensus,'' \emph{arXiv preprint arXiv:2101.03300}, 2021.

\bibitem{lu2019blockchain}
Y.~Lu, X.~Huang, Y.~Dai, S.~Maharjan, and Y.~Zhang, ``Blockchain and federated learning for privacy-preserved data sharing in industrial iot,'' \emph{IEEE Transactions on Industrial Informatics}, vol.~16, no.~6, pp. 4177--4186, 2019.

\bibitem{qu2021proof}
X.~Qu, S.~Wang, Q.~Hu, and X.~Cheng, ``Proof of federated learning: A novel energy-recycling consensus algorithm,'' \emph{IEEE Transactions on Parallel and Distributed Systems}, vol.~32, no.~8, pp. 2074--2085, 2021.

\bibitem{li2020federated}
T.~Li, A.~K. Sahu, M.~Zaheer, M.~Sanjabi, A.~Talwalkar, and V.~Smith, ``Federated optimization in heterogeneous networks,'' \emph{Proceedings of Machine learning and systems}, vol.~2, pp. 429--450, 2020.

\bibitem{wang2020tackling}
J.~Wang, Q.~Liu, H.~Liang, G.~Joshi, and H.~V. Poor, ``Tackling the objective inconsistency problem in heterogeneous federated optimization,'' \emph{Advances in neural information processing systems}, vol.~33, pp. 7611--7623, 2020.

\bibitem{li2020blockchain}
Y.~Li, C.~Chen, N.~Liu, H.~Huang, Z.~Zheng, and Q.~Yan, ``A blockchain-based decentralized federated learning framework with committee consensus,'' \emph{IEEE Network}, vol.~35, no.~1, pp. 234--241, 2020.

\bibitem{liu2020fedcoin}
Y.~Liu, Z.~Ai, S.~Sun, S.~Zhang, Z.~Liu, and H.~Yu, ``Fedcoin: A peer-to-peer payment system for federated learning,'' in \emph{Federated learning: privacy and incentive}.\hskip 1em plus 0.5em minus 0.4em\relax Springer, 2020, pp. 125--138.

\bibitem{sokhankhosh2024proof}
A.~Sokhankhosh and S.~Rouhani, ``Proof-of-collaborative-learning: A multi-winner federated learning consensus algorithm,'' in \emph{IEEE International Conference on Blockchain (Blockchain)}.\hskip 1em plus 0.5em minus 0.4em\relax IEEE, 2024, pp. 370--377.

\bibitem{benet2014ipfs}
J.~Benet, ``Ipfs-content addressed, versioned, p2p file system,'' \emph{arXiv preprint arXiv:1407.3561}, 2014.

\bibitem{jacob2018quantization}
B.~Jacob \emph{et~al.}, ``Quantization and training of neural networks for efficient integer-arithmetic-only inference,'' in \emph{Proceedings of the IEEE conference on computer vision and pattern recognition}, 2018, pp. 2704--2713.

\bibitem{xing2023zero}
Z.~Xing \emph{et~al.}, ``Zero-knowledge proof-based practical federated learning on blockchain,'' \emph{arXiv preprint arXiv:2304.05590}, 2023.

\bibitem{flock2024mainnet}
\BIBentryALTinterwordspacing
{FLock Team}, ``Flock: Federated machine learning on blockchain,'' FLock, Whitepaper, 2024. [Online]. Available: \url{https://www.flock.io/whitepaper}
\BIBentrySTDinterwordspacing

\bibitem{asi2024intelligence}
\BIBentryALTinterwordspacing
{SingularityNET}, {Fetch.ai}, and {Ocean Protocol}, ``Artificial superintelligence (asi) alliance vision paper: Building decentralized artificial superintelligence,'' SingularityNET, Fetch.ai, Ocean Protocol, Tech. Rep., 2024. [Online]. Available: \url{https://docs.superintelligence.io/artificial-superintelligence-alliance/vision-paper}
\BIBentrySTDinterwordspacing

\bibitem{heyndrickx2023melloddy}
W.~Heyndrickx \emph{et~al.}, ``Melloddy: Cross-pharma federated learning at unprecedented scale unlocks benefits in qsar without compromising proprietary information,'' \emph{Journal of chemical information and modeling}, vol.~64, no.~7, pp. 2331--2344, 2023.

\bibitem{behera2022fedsyn}
M.~R. Behera, S.~Upadhyay, S.~Shetty, S.~Priyadarshini, P.~Patel, and K.~F. Lee, ``Fedsyn: Synthetic data generation using federated learning,'' \emph{arXiv preprint arXiv:2203.05931}, 2022.

\bibitem{wrona2018does}
K.~Wrona and M.~Jarosz, ``Does nato need a blockchain?'' in \emph{MILCOM 2018-2018 IEEE Military Communications Conference (MILCOM)}.\hskip 1em plus 0.5em minus 0.4em\relax IEEE, 2018, pp. 667--672.

\bibitem{agriflchain2025}
T.~Manoj, K.~Makkithaya, and V.~G. Narendra, ``A blockchain-assisted trusted federated learning for smart agriculture,'' \emph{SN Comput. Sci.}, vol.~6, p. 221, 2025.

\end{thebibliography}

\end{document}